\documentclass[12pt]{iopart}
\newtheorem{lemma}{Lemma}
\newtheorem{proposition}{Proposition}
\newtheorem{theorem}{Theorem}

\newtheorem{definition}{Definition}

\usepackage{iopams}
\usepackage{graphicx,amssymb}
\usepackage{color}

\usepackage{ifthen}

\usepackage[latin1]{inputenc}
\usepackage{amsfonts}

\begin{document}

\title[Positioning systems in Minkowski space-time]
{Positioning systems in Minkowski space-time: from emission to inertial
coordinates}

\author{Bartolom\'{e} Coll, Joan Josep Ferrando and \\
Juan Antonio Morales-Lladosa}

\address{Departament d'Astronomia
i Astrof\'{\i}sica, Universitat de Val\`encia, \\E-46100 Burjassot,
Val\`encia, Spain.}

\ead{bartolome.coll@uv.es; joan.ferrando@uv.es;
antonio.morales@uv.es}

\begin{abstract}
The coordinate transformation between emission coordinates and
inertial coordinates in Minkowski space-time is obtained for
arbitrary configurations of the emitters. It appears that a
positioning system always generates two different coordinate
domains, namely, the front and the back emission coordinate domains.
For both domains, the corresponding covariant expression of the
transformation is explicitly given in terms of the emitter
world-lines. This task requires the notion of orientation of an
emitter configuration. The orientation is shown to be computable
from the emission coordinates for the users of a `central' region of
the front emission coordinate domain. Other space-time regions
associated with the emission coordinates are also outlined.
\end{abstract}




\section{Introduction}
\label{intro} A relativistic positioning system consists of a set of
four clocks $A$ ($A=1,2,3,4$) broadcasting their respective proper times
$\tau^A$ by means of electromagnetic signals. Then, every event reached
by the signals is naturally labelled by
the four times $\{\tau^A\}$: the emission coordinates of this event.
Elsewhere \cite{white}, we have presented a brief report on
relativistic positioning and related issues, providing a background
with current references on the subject.

A user of a positioning system that receives the four times
$\{\tau^A\}$ knows his own coordinates in the emission system. Then,
if he wants to know his position in another coordinate system, he
must obtain the transformation between both coordinate systems.
Thus, we must solve the following important problem in relativistic
positioning. Suppose that the world-lines of the emitters
$\gamma_A(\tau^A)$ are known in a coordinate system
$\{x^{\alpha}\}$: can the user obtain his coordinates in this system
if he knows his emission coordinates $\{\tau^A\}$? Or, slightly more
general, can the coordinate transformation
$x^{\alpha}=\kappa^{\alpha}(\tau^A)$ be obtained?

The main goal of this paper is to solve this question in Minkowski
space-time or, more explicitly, to obtain the coordinate
transformation between emission and inertial coordinates for
arbitrary world-lines of the emitters.

In a two-dimensional approach to relativistic positioning systems
this query is rather simple. Indeed, in this case the knowledge of
the emitter's world-lines in a chosen null coordinate system gives
the coordinate transformation between this null system and the
emission coordinates at once \cite{D2a, D2b}. This fact facilitates
an analytical study of the positioning systems defined by inertial
emitters in Minkowski plane \cite{D2a}, and those defined by
stationary emitters in both Minkowski and Schwarzschild planes
\cite{D2b}.

The general properties of the emission coordinates for the
four-dimensional case have been analyzed in \cite{D4a}.
Nevertheless, in this generic case it is difficult to solve the
above stated problem. The study of specific three-dimensional
situations \cite{Pozo-Escola} has shed light on the complex geometry
of the domains and codomains of the emission coordinates. On the
other hand, the transformation between emission and inertial
coordinates has recently been obtained for emitters following
particular inertial motions in flat space-time, and also considering
the immediate vicinity of a Fermi observer in the Schwarzschild
geometry \cite{BiniMas}.

Here we present the solution to this problem for a generic
configuration of the emitters in Minkowski space-time. We show that
two emission domains exist which are called the front emission
coordinate domain and the back emission coordinate domain, and we
give the coordinate transformation for each one. The transformation
$x^{\alpha}=\kappa^{\alpha}(\tau^A)$ between inertial and emission
coordinates is given in a covariant way in terms of the world-lines
$\gamma_A(\tau^A)$ of the emitters. The compact covariant expression
of our result is a powerful tool for use in subsequent applications:
to obtain the coordinate transformation for specific configurations
of the emitters or, under perturbation methods, modeling more
realistic gravitational situations.

The paper is structured as follows. In section \ref{sec-2}, we pose
the problem to be solved, defining the emission regions and the main
relations governing the coordinate transformation: the null
propagation conditions and the emission conditions. By choosing one
of the emitters arbitrarily as reference emitter, in section
\ref{sec-3}, we show that the null propagation equations of a
positioning system are equivalent to a rank-three linear system and
a sole quadratic equation. Then, section \ref{sec-4} is devoted to
obtaining the general solution of the linear system, and in section
\ref{sec-5} we impose the remaining quadratic equation and we obtain
a general form for both solutions. In section \ref{sec-6}, we apply
the emission conditions which guarantee that the solutions are
physically interpretable as emission solutions. In section
\ref{sec-7} we define the orientation of the positioning system with
respect to an event, a concept which allows us to give the main
result of this paper in compact form, namely, the  explicit
expression of the coordinate transformation between emission and
inertial coordinates. The computational and/or observational
determination of the orientation of the positioning system is
analyzed in section \ref{sec-8}. Section \ref{sec:discussion} deals
with the analysis of our results and comments about ongoing work in
progress and on further practical applications. Finally, an appendix
is devoted to presenting a technical proof of the results given in
section \ref{sec-4}.

A short communication on this work has been presented at the Spanish
Relativity meeting ERE-2008 \cite{NosERE08}.

\section{Statement of the problem. Emission regions and emission relations}
\label{sec-2}

\subsection{The emission region ${\cal R}$ of a positioning system}
\label{sec-2-1}

    Let us consider a {\em positioning system} in Minkowski space-time
${\cal M}^4$, and let $\gamma_A(\tau^A)$, $A = 1, \ldots, 4$, be the
world-lines of its four distinct emitters $A$ (clocks) watching
their proper time $\tau^A$. These four times, broadcast by means of
electromagnetic signals, will reach some region, say the {\em
emission region} ${\cal R}$, of events of ${\cal M}^4$. In the interior of
${\cal R}$, the four times $\{\tau^A\}$ converging at every event define
generically the so called {\em emission coordinates}
\cite{D2a,D2b,D4a,Coll-ERE-2000,ToloERE05}.

    It is clear that an event {\small P} of ${\cal M}^4$ belongs to the
emission region ${\cal R}$ if and only if there exists a past
directed null geodesic from {\small P} to every emitter
$\gamma_A(\tau^A)$ for some value of $\tau^A$. For future
discussions, we need to make the ingredients of this assertion more
explicit.

    Let us denote by $x$ the position vector of {\small P} with respect
to the origin {\small O} of some inertial chart $\{x^{\alpha}\}$, $x
\equiv$ {\small OP}, and by $\gamma_A$ the position vector of the
emitters with respect to this chart, $\gamma_A \equiv {\rm{\small
O}} \gamma_A(\tau^A)$.  Then, in order for {\small P} to belong to
${\cal R}$, the four vectors
\begin{equation} \label{defellnull}
    m_A \equiv x - \gamma_A \ ,
\end{equation}
which represent the trajectories followed by the electromagnetic
signals in vacuum issued from the emitters $A$ (see Fig.
\ref{Fig-OneTwoEvent&Clocks}a), must verify the {\em null
propagation conditions} \textsf{L}\,:
\begin{equation*}
{\rm \textsf{L}:}\quad   (m_A)^2 =  0 \ , \quad \forall A \ .
\end{equation*}
Furthermore, these four vectors have to be future-pointing or, in other
words, must verify the {\em emission conditions} {\rm \textsf{E}}\,:
\begin{equation*}
{\rm \textsf{E}:}\quad     \epsilon\,u \cdot m_A  < 0 \ , \quad
\forall A \ ,
\end{equation*}
where $u$ is any given, everywhere non vanishing, time-like vector
field defining the arrow of time and $2\epsilon$, $\epsilon = \pm
1$, is the metric signature.

The events $x$ where the emission condition \textsf{E} holds must be
{\em receivers}. But it is worth noting that the null propagation
conditions \textsf{L} also apply in the case of `active' events $x$,
able to send or to reflect null signals to one or more of the
emitters.
Such other {\em location systems}, as the physical
realizations of coordinate systems are called here, have been dealt with elsewhere
\cite{Ehlers+, GraviMas, Perlick} but will not be considered here.

Nevertheless, in obtaining our coordinate transformations, we will
need to consider, besides the emission condition \textsf{E}, its
`causally dual' {\em reception conditions} {\rm \textsf{R}} \,:
\begin{equation*} 
{\rm \textsf{R}:} \quad \quad \epsilon \, u \cdot m_A  > 0  \ , \quad
\forall A \ ,
\end{equation*}
both condensed in the {\em emission-reception conditions} {\rm \textsf{E-R}}\,:
\begin{equation*} 
{\rm \textsf{E-R}:} \quad \epsilon \, m_A \cdot m_B < 0 \ , \quad
\forall A\, , B \ ,
\end{equation*}
as it is easy to argue.

    Now, the above assertion about {\small P} and ${\cal R}$ may be stated as
follows: {\em an event} {\small P} {\em belongs to the emission
region ${\cal R}$ of a positioning system of emitters $\gamma_A(\tau^A)$ if,
and only if, its position vector $x$ in
some inertial chart verifies the null and emission conditions {\rm
\textsf{L}} and  {\rm \textsf{E}}, respectively, for some values of
the proper times $\tau^A$ of the emitters.}

\subsection{The characteristic emission function $\Theta$ of a positioning
system and the emission coordinate region $\mathcal{C}$ of the space-time} \label{sec-2-2}

    It then turns out that a positioning system may be considered as a
device $\Theta$ that physically associates, to every event of the
emission region ${\cal R}$, a set of four times $\{\tau^A\}$.

    From a formal point of view, this device is nothing but an application
$\Theta :$ ${\cal R} \longrightarrow \mathcal{T}$, henceforth called
the {\em characteristic emission function} of the positioning
system, that, with every event $x$ of the emission region ${\cal R}$
of the space-time, associates four times $\tau^A$ of the {\em grid}
$\mathcal{T}$ of the $\tau's$, $\Theta :$ $x \longmapsto$ $(\tau^A)
=$ $\Theta(x)$.

    The grid $\mathcal{T}$ is nothing but the Cartesian product $\mathcal{T} \equiv$
$\stackrel{4}{\times}\{{\tau}\}$ $\approx$ $\mathbb{R}^4$ of the spaces
(real lines) of the variables $\tau$ (for details on the concept and role
of the grid, see for example \cite{D2a, D2b, Pozo-Escola}).

In this grid $\mathcal{T}$, the image $\Theta({\cal R})$ of the emission
region ${\cal R}$ by the characteristic emission function $\Theta$,
is called the {\em emission co-region} of the positioning system,
and is denoted by $^\Theta\!{\cal R}$, $^\Theta\!{\cal R} = \Theta({\cal R})$.
The points of this region of
the grid $\mathcal{T}$ are the quadruplets of times that can really be
received in the space-time, so that they are related to the
space-time events at which they are measured. In this sense, this
emission co-region $^\Theta\!{\cal R}$ is the sole region of the
grid $\mathcal{T}$ which possesses a physical meaning (the other quadruplets
of $\mathcal{T}$ are a convenient mathematical completion of the emission
co-region but with {\em no} relation to space-time events).

    Obviously, if the world-lines of the four emitters are sufficiently
smooth and broadcast their proper time continuously, the emission co-region
$^\Theta\!{\cal R} \subset$ $\mathcal{T}$ is connected and, because of
the regularity of the light cones in Minkowski space-time, ${\cal R}$ is
connected too. But this property does not guarantee that the emission
function $\Theta$, $\tau^A =$ $\Theta^A(x)$,
is invertible in ${\cal R}$.

    For $\Theta$ to be invertible, the gradients $d\tau^A$, normal to
the hypersurfaces $\tau^A =$ constant, have to be {\em well defined}
and {\em linearly independent}. But, because in Minkowski space-time
our light cones are everywhere differentiable up to on their
vertices, i.e. on the emitter world-lines $\gamma_A$, the gradients
$d\tau^A$ are well defined everywhere in ${\cal R}$ up to on ${\cal
R} \cap \left(\cup_A\gamma_A(\tau^A)\right)$. Then, on the region
${\cal R} - \left(\cup_A\gamma_A(\tau^A)\right)$, because the
$d\tau^A$ are metrically collinear to the above null vectors $m_A$,
the linear independence of the $d\tau^A$ may be expressed by the
{\em coordinate condition} \textsf{C}:
\begin{equation*} 
{\rm \textsf{C}:}\quad      m_1 \wedge m_2 \wedge m_3 \wedge m_4
\neq 0 \ .
\end{equation*}
This condition \textsf{C} is equivalent to say that $j_\Theta(x)
\not=0$, where $j_\Theta(x)$ is the determinant of the Jacobian
matrix $J_\Theta(x)$ of $\Theta$, so that the locus where the
$d\tau^A$ are linearly dependent is the hypersurface ${\mathcal{J}}$
in ${\cal R}$ of equation ${\mathcal{J}} \equiv \{j_\Theta(x) =
0\}$. We shall call the regions ${\mathcal{D}} \equiv {\mathcal{J}}
\cup (\cup_A \gamma_A(\tau^A))$ and $\mathcal{C} \equiv {\cal R} -
\mathcal{D}$ the {\em emission degenerate region}, and the {\em
emission coordinate region} respectively, ${\cal R} = {\cal C} \cup
{\cal D}$.

It is to be noted that, the condition \textsf{C} being satisfied on
its events, $\mathcal{C}$ is an open set. For the same reason,
$\Theta$ is invertible in $\mathcal{C}$, so that all its events may
be locally labeled by the coordinates $\{\tau^A\}$. Nevertheless,
this does not means necessarily that $\mathcal{C}$ be a {\em
coordinate domain} of a local chart $(\mathcal{C}, \Theta)$,
because the condition \textsf{C} assures only the {\em local}
invertibility of $\Theta$. In fact, we will see that $\mathcal{C}$
is {\em not} a coordinate domain, but the union of {\em two}
coordinate domains; this is why we have called $\mathcal{C}$ the
emission coordinate {\em region}. The region in the grid
$\mathcal{T}$ where the characteristic emission function $\Theta$,
$\tau^A =$ $\Theta^A(x)$, is locally invertible is the region
$^\Theta\mathcal{C} \equiv$ $\Theta(\mathcal{C})$, and will be
called the {\em emission coordinate co-region} of the grid
$\mathcal{T}$.

\subsection{The main relations of a positioning system}
\label{sec-2-3}

When the world-lines of the emitters are known, the {\em main
relations} of a positioning system are the set $\{{\rm \textsf{L}}
, {\rm \textsf{E}} \}$ of the null propagation conditions {\rm
\textsf{L}} and the emission conditions {\rm \textsf{E}}.

In terms of the position vectors $\gamma_A$ of these world-lines
$\gamma_A(\tau^A)$, they give rise, by (\ref{defellnull}), to the
{\em null propagation equations}
\begin{equation} \label{ellnull}
{\rm \textsf{L}:}\quad    (x - \gamma_A)\cdot(x - \gamma_A) = 0 \ ,
\quad \forall A \ .
\end{equation}
and to the {\em emission inequalities}
\begin{equation} \label{futurepointing}
{\rm \textsf{E}:}\quad    \epsilon\,u \cdot(x - \gamma_A)  < 0     \
, \quad \forall A \ ,
\end{equation}
where $u$ is any given, everywhere non vanishing, future-pointing
time-like vector.

To invert the function $\Theta$ in the emission coordinate co-region
$^\Theta\mathcal{C}$ is to solve the main relations (\ref{ellnull}),
 (\ref{futurepointing}) in $x$, $x = \kappa(\tau^A)$, for values of
 the $\tau^A$'s verifying the coordinate condition \textsf{C} or,
 by (\ref{defellnull}), the {\em coordinate inequality}:
\begin{equation} \label{coordcondit}
{\rm \textsf{C}:}\quad      (x - \gamma_1)  \wedge (x - \gamma_2)
 \wedge (x - \gamma_3) \wedge (x - \gamma_4) \neq 0 \ .
\end{equation}

These solutions $x = \kappa(\tau^A)$ may alternatively be read as
the position vectors of the events whose past light cone cuts the
emitter world-lines $\gamma_A(\tau^A)$ at their times $\tau^A$. The
components $x^\alpha = \kappa^\alpha(\tau^A)$ of $x$ then define
the coordinate transformation between the emission coordinates
$\{\tau^A\}$ and the inertial ones $\{x^\alpha\}$.

The main object of this paper is to obtain this coordinate
transformation.

\section{The null propagation equations \textsf{L}}
\label{sec-3}

    When $\{\tau^A\}$ are the emission coordinates of the event $x$, the
emitters are at the events $\{\gamma_A(\tau^A)\}$. These four events
define the internal {\em configuration} of the emitters for the
event $x$.

    One can try to solve the null propagation equations (\ref{ellnull}) in
$x$, $x = \kappa(\tau^A)$ directly, but it is better to first
carefully separate the ingredients intrinsically related to the
configuration of the four emitters $A$, which are independent of the
inertial chart $\{x^\alpha\}$, from the ingredients related to this
chart, which is independent of the configuration of the emitters.

\begin{figure}
\centerline{
\parbox[c]{0.65\textwidth}{\includegraphics[width=0.80\textwidth]{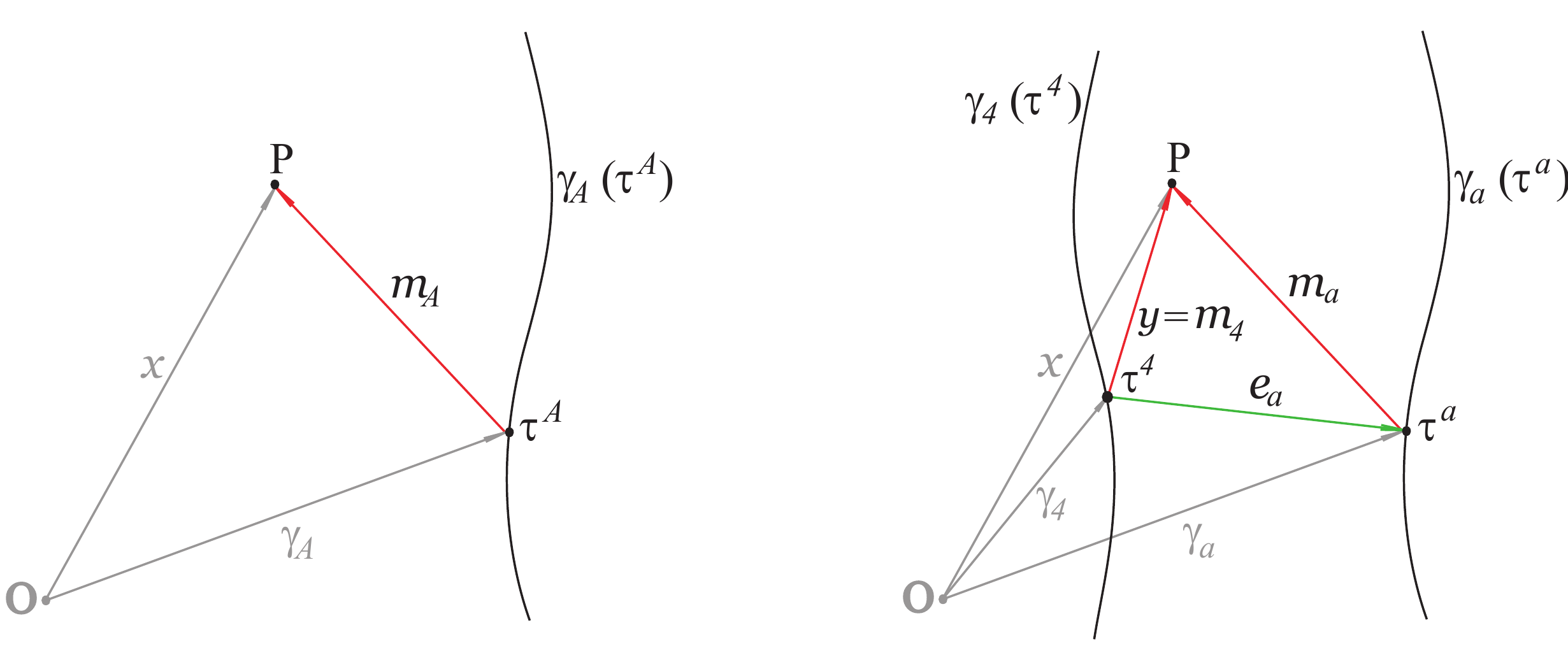}}}
\caption{(a) The light-like vectors $m_A \equiv x - \gamma_A$
represent the trajectories followed by the electromagnetic signals
between the emitters and an event P in the emission region. $x$ is
the position vector of this event with respect to some inertial
chart, and $\gamma_A$, $A=1,2,3,4$, are the position vectors of the
emitters with respect to this chart. (b) If we choose the emitter
$4$ as origin (reference emitter), the relative positions of the others (referred
emitters) are $e_a = \gamma_a - \gamma_4$, $a=1,2,3$, and the
position vector of the event P is $y = m_4$. \label{Fig-OneTwoEvent&Clocks}}
\end{figure}

    For this purpose, we shall arbitrarily choose one of the emitters, say
$A = 4$, as {\em reference emitter} and we shall search the solution $x$
under the form:
\begin{equation} \label{xgamma4y}
    x = \gamma_4 + y \ ,
\end{equation}
where $y$ is the solution to the null propagation equations
(\ref{ellnull}) when the origin is chosen at the position $\gamma_4$
of the emitter $4$ (see Fig. \ref{Fig-OneTwoEvent&Clocks}b). Thus, $y$ is submitted
to
\begin{equation} \label{ellnullwithy}
{\rm \textsf{L}_4:} \quad      (y - e_A)\cdot(y - e_A) = 0   \ ,
\quad \forall A \  ,
\end{equation}
where
\begin{equation} \label{defeA}
    e_A = \gamma_A - \gamma_4 \ .
\end{equation}
Now, splitting these relations for $A = 4$ and $A = a = 1,2, 3$, we have
\begin{equation}  \label{defe4ea}
    e_4 = 0 \ , \quad  e_A = e_a
\end{equation}
and, for the null propagation equations (\ref{ellnullwithy}),
\begin{equation} \label{ellnullwithy4a}
    \begin{array}{ll}
    y^2 = 0 \ ,  & A = 4 \ ,\\
    y^2 - 2\,e_a\cdot y + (e_a)^2 = 0 \ , & A = a \ .
    \end{array}
\end{equation}
Noting that $e_a$ are the relative position vectors of the
{\em referred} emitter $A = a$ with respect to the reference emitter
$A = 4$, and that the half of their squares, represent the {\em world
functions} \cite{Synge} of the referred and reference emitters, we have:
%
%
%
%
\begin{lemma} \label{lem-decommaineqs}
The null propagation equations {\rm (\ref{ellnullwithy})} are equivalent
to the {\em main linear system}
\begin{equation} \label{linearsys}
    e_a \cdot y  = \Omega_a  \ , \quad a = 1,2,3  \ ,
\end{equation}
and the {\em main quadratic equation}
\begin{equation}\label{quadratEq}
y^2 = 0  \ ,
\end{equation}
where the scalars $\Omega_a$, world functions of the endpoints of the
vectors $e_a$, are given by
\begin{equation} \label{squareea}
\Omega_a \equiv \frac{1}{2} (e_a)^2 \ .
\end{equation}
\end{lemma}
%
%
%

In the study of the null propagation equations we are supposing that
the emitted times $\{\tau^A\}$ received at $x$ constitute the
emission coordinates of $x$ so that the coordinate inequality
(\ref{coordcondit}) fulfils. From the definitions (\ref{defeA})
this condition writes
\begin{equation} \label{C-4}
{\rm \textsf{C}_4:} \quad    \chi \cdot y \neq 0 \ ,
\end{equation}
where $\chi$ is the \emph{configuration vector} defined by
\begin{equation} \label{defvectorchi}
\chi \equiv * (e_1 \wedge e_2 \wedge e_3)  \ ,
\end{equation}
with $*$ being the Hodge dual operator. Note that, under the form
${\rm \textsf{C}_4}$ of ${\rm \textsf{C}}$, $\chi$ is necessarily
non vanishing:
\begin{equation} \label{nonvanishchi}
\chi \neq 0 \ ,
\end{equation}
which means that the configuration
$\{\gamma_A(\tau^A)\}$ determines a hyperplane. This {\em configuration
hyperplane} contains the four events $\{\gamma_A(\tau^A)\}$ and is orthogonal
to the configuration vector $\chi$. The sign of $\chi^2$ provides
the causal character (space-like, light-like or time-like) of this
hyperplane. Thus, in the emission coordinate region $\mathcal{C}$ one
can distinguish three disjoint regions:
\begin{description}
\item
the {\em space-like configuration
region} $\quad \mathcal{C}_s \equiv \{ x \in \mathcal{C} \ | \ \epsilon \,
\chi^2 < 0\}$,
\item
the {\em null configuration region} $\  \qquad \quad
\mathcal{C}_\ell \equiv \{ x \in \mathcal{C} \ | \ \ \, \chi^2 =
0\}$,
\item
the {\em time-like configuration region} $\ \quad \mathcal{C}_t \equiv \{ x \in
\mathcal{C} \ | \ \epsilon \, \chi^2 > 0\}$.
\end{description}
with $\mathcal{C} = \mathcal{C}_s \cup \mathcal{C}_\ell \cup \mathcal{C} _t$.

The ingredients $\Omega_a$ and $\chi$ of the internal configuration
of the emitters make use of the relative position vectors $e_a$ and
their double exterior product $e_1 \wedge e_2 \wedge e_3$. Let us
complete them by introducing their simple exterior products in the form
\begin{equation}\label{defEa}
    E^a \equiv  *(e_{a+1} \wedge e_{a+2}) \ ,
\end{equation}
where the operations on the indices obviously have to be understood
modulo 3.

In looking for the solutions to the null propagation equations
(\ref{ellnullwithy}), it is convenient to first start solving the
main linear system (\ref{linearsys}) in $y$ and then constraint it
to verify the main quadratic equation (\ref{quadratEq}). Then, the
solution(s) $y = \iota(\tau^A)$ so obtained, incremented by the
position $\gamma_4$ of the reference emitter $A = 4$, will give by
(\ref{xgamma4y}) the solution(s) to the null propagation equations
(\ref{ellnull}), $x = \kappa(\tau^A)= \gamma_4 + \iota(\tau^A)$,
i.e. the wanted coordinate transformation between emission
coordinates and inertial ones. The next section is devoted to
solving the main linear system (\ref{linearsys}).


\section{The main linear system}
\label{sec-4}

The regularity condition (\ref{nonvanishchi}) for the configuration
vector $\chi$ means that the rank of the main linear system
(\ref{linearsys}) is exactly 3. As a consequence, its general
solution $y$ depends on a sole parameter, say $\lambda$, so that
denoting by $y_*$ a particular solution, one has $y = y_* + \lambda
\chi$ because, by (\ref{defvectorchi}), $\chi$ is a vector such that
$e_a\cdot \chi = 0$.
We can thus state:
%
%
%
\begin{lemma}\label{Y=chi}
In the emission coordinate region $\mathcal{C}$, the general solution to the main
linear system {\rm (\ref{linearsys})} is of the form
\begin{equation}\label{solutyy*}
    y = y_* + \lambda\chi \ ,
\end{equation}
where the parameter $\lambda$ takes arbitrary values, $\chi$ is the
configuration vector {\rm (\ref{defvectorchi})} and $y_*$ is a particular solution.
\end{lemma}
%
%

A tentative study of explicit expressions of the particular solution
$y_*$ shows that, for every emission coordinate domain, {\em there
is not a sole analytical function, depending exclusively of the
elements of the configuration $\{\gamma_A(\tau^A)\}$ of the
emitters, that is valid throughout the domain}. This is why it is
{\em necessary} to introduce an external element to obtain a sole
analytical expression. We shall see that it is {\em sufficient} for
this external element to be a vector field $\xi$ {\em transversal}
to the configuration, i.e. such that
\begin{equation}\label{deftransversal}
    \xi \cdot \chi \neq 0 \ ,
\end{equation}
and otherwise arbitrary. Denoting by $i(\ )$ the interior product,
it may then be shown (see \ref{A-SolCov} for the proof):

%
%
%
\begin{proposition}\label{expresSolPart}
In all the emission coordinate region $\mathcal{C}$, the general
solution to the main linear system {\rm (\ref{linearsys})},
\begin{equation*}
    e_a \cdot y  = \Omega_a  \ , \quad a = 1,2,3  \  ,
\end{equation*}
is of the form
\begin{equation}\label{solutyybot}
    y = y_* + \lambda\chi \ ,
\end{equation}
where the parameter $\lambda$ takes arbitrary values, $\chi$
is the configuration vector
\begin{equation*}
 \chi \equiv *(e_1 \wedge e_2 \wedge e_3) \ ,
\end{equation*}
and $y_*$ is the particular solution orthogonal to a chosen
transversal vector $\xi$, $\xi \cdot \chi \neq 0$, given by
\begin{equation}\label{solutyorto}
y_* \equiv \frac{1}{\xi\cdot\chi} \ i(\xi)H \ ,
\end{equation}
where the bivector $H$ is, like the vector $\chi$, a function of the
configuration of the emitters,
\begin{equation}\label{defschiHEa}
 H \equiv  \Omega_a E^a \ , \quad E^a \equiv *(e_{a+1} \wedge e_{a+2})
\ .
\end{equation}
\end{proposition}
%

    The direct and simple discussion that has enabled us to reach
proposition~\ref{expresSolPart} is the result of a careful search
in order to find, in every coordinate domain, a {\em sole}, {\em
general} and {\em covariant} expression for the coordinate
transformation between emission coordinates $\{\tau^A\}$ and
inertial ones $\{x^\alpha\}$.

    The last two requirements of generality and covariance enable us
to find {\em once and for all} and {\em for any inertial system}
(and in {\em compact form} due to the intrinsic vector formalism)
the coordinate transformation in question. But they do not suffice
to lead by themselves to a sole expression valid for all the
configurations, i.e. valid at {\em all} the events of every emission
coordinate domain. For example, if we were to work directly on the
main linear system in the cases of a configuration generating a
nonnull vector $\chi$, we would find the solution
\begin{equation} \label{xiequalchi}
y =  \frac{1}{\chi^2}\  i(\chi)H + \lambda\chi \ ,
\end{equation}
as now directly follows from (\ref{solutyorto}) and
(\ref{solutyybot}) for the choice $\xi$ $=\chi$. But, meanwhile in fact
there is neither physical nor mathematical reason for the solution
to lose analyticity, this expression
(\ref{xiequalchi}) becomes manifestly undefined on the null
configurations, $\chi^2 = 0$. This is why it is {\em necessary} to
introduce an external element to find the sole regularized analytical
expression above mentioned.

    This external element, the transversal vector field $\xi$, independent
of the con\-fi\-guration of the emitters, is otherwise
arbitrary. But, meanwhile due to lemma \ref{Y=chi} we know that the
main linear system depends on a sole parameter $\lambda$, the
expressions (\ref{solutyybot}), (\ref{solutyorto}) of $y$ apparently
seem to indicate that it depends on the $(4+1=)$ 5  arbitrary
parameters $\{\xi;\lambda\}$. Thus, what is the precise role played
by this transversal field $\xi$?

    From the definition (\ref{defschiHEa}) of the bivector
$H$ it is clear that its dual $*H$ is orthogonal to $\chi$,
$i(\chi)*H = 0$, so that, one has $H$ $= \chi \wedge a$ for some
vector $a$. With this expression of $H$ it is easy to prove from
(\ref{solutyybot}) and (\ref{solutyorto}) that the variation of the
solution $y$ with respect to the parameters $\xi$ verifies
\begin{equation} \label{partialyrespxi}
\frac{\partial y}{\partial \xi} \wedge \chi = 0  \ ,
\end{equation}
which shows that the changes in $\xi$ may be absorbed by $\lambda$, i.e.
that the parameters $\xi^\alpha$ are not {\em essential} in the Eisenhart
sense \cite{Eisenhart}.
Nevertheless their elimination gives rise to different expressions
for the causally different regions that one can find in the whole
emission coordinate region.


\section{The main quadratic equation}
\label{sec-5}

Obtaining the transformation $x^\alpha = \kappa^\alpha(\tau^A)$ between
emission coordinates $\{\tau^A\}$ and inertial ones $\{x^\alpha\}$
amounts to determine the intersection of the four future light cones
emitted by the emitters at the configuration
$\{\gamma_A(\tau^A)\}$ or, in the dual interpretation, to determine the
vertex of the past light cone that contains the emitters in such a
configuration.

    According to proposition \ref{expresSolPart}, the solutions $y$ of the
main linear system ($\ref{linearsys}$) depend linearly of the
parameter $\lambda$, and consequently,    describe a straight line
in ${\cal M}^4$. The vertex that we are searching for is
consequently on this straight line and, according to our choice of
reference emitter, is the point on it related to the emitter $A=4$
by the null vector $m_4 $ $= x - \gamma_4$ $ = y$. This is nothing
more than what the main quadratic equation (\ref{quadratEq})
expresses: $y^2 = 0$. This cone, whose vertex is the emitter $A =
4$, cuts the straight line (\ref{solutyybot}) at a value of
$\lambda$ given implicitly by:
\begin{equation}\label{firstquadratic}
\chi^2 \lambda^2 + 2(\chi\cdot y_*)\lambda + y_*^2 = 0  \ ,
\end{equation}
and it is this value of $\lambda$ which, put in the expression
(\ref{solutyybot}) of proposition \ref{expresSolPart}, enables us to
obtain the coordinate transformation $y = \iota(\tau^A)$ that we are
searching for.

But, before obtaining it, and in order to be sure of its real
character, it is convenient to observe that the sole assumption we
have made so far to solve the null propagation equations is that of
the regularity of the configuration data $\{\gamma_A(\tau^A)\},$
i.e. of the non-vanishing of the configuration vector $\chi,$ $\chi
\neq 0.$ This assumption is only a \emph{strict} part of the
coordinate condition $\textsf{C}$, as its form $\textsf{C}_4$, $\chi
\cdot y \neq 0$, shows.

The study of degenerate configuration data $\chi = 0$ and of the
degenerate coordinate condition $\chi\cdot y = 0$ is interesting to
gain more in-depth knowledge of the physical conditions leading to
their presence, in order to avoid, control or predict them. But we
shall restrict ourselves here, as already stated, to obtaining the
coordinate transformation between emission coordinates $\{\tau^A\}$
and inertial ones $\{x^\alpha\}.$ For this purpose, from now on, we
shall work under the coordinate condition $\textsf{C}.$

In fact, it is this condition that guarantees the real character of
the solutions of the null propagation system. To demonstrate this,
we shall separately analyze the cases of null and non null
configurations.

\vspace{2mm} \noindent {\textbf{Null configurations.}} In the region
where the configuration vector is null, $\chi^2 = 0,$ expression
(\ref{solutyybot}) for the solution $y,$ $y = y_* + \lambda\chi,$
leads to $\chi \cdot y = \chi \cdot y_*,$ so that the last
expression (\ref{C-4}) for the coordinate condition $\textsf{C}$
ensures the well defined character of $\lambda$ in
(\ref{firstquadratic}),
\begin{equation}\label{solelambda}
\lambda = - \frac{y_*^2}{2(\chi\cdot y_*)} \, ,
\end{equation}
and we have
%
%
\begin{proposition}\label{yparachidosnulo}
Under the coordinate condition {\rm \textsf{C}}, for null
configurations, $\chi^2 = 0,$  the null propagation equations {\rm
(\ref{ellnullwithy})} admit a real and single solution $y$ given by:
\begin{equation}\label{yunica}
    y = y_* - \frac{y_*^2}{2(\chi\cdot y_*)}\, \chi  \, ,
\end{equation}
where $y_*$ and $\chi$ are respectively given by {\rm
(\ref{solutyorto})} and {\rm (\ref{defvectorchi})}.
\end{proposition}
%
%
%

\noindent {\textbf{Non-null configurations.}} In the regions
where the configuration vector is non-null, $\chi^2 \neq 0,$
(\ref{firstquadratic}) gives:
\begin{equation}\label{standardexpress}
\lambda_\pm = \frac{1}{\chi^2}\left( - \chi\cdot y_* \pm
\sqrt{\Delta} \right) \ , \quad \Delta \equiv (\chi\cdot y_*)^2 -
\chi^2\,y_*^2 \ .
\end{equation}
But, for these non-null configurations, we can choose $\xi = \chi$
in the expression (\ref{solutyorto}) of the particular solution
$y_*$ to the main linear system (\ref{linearsys}). Let us denote by
$c$ the corresponding $y_*$ :
\begin{equation}\label{defc}
    c =  \frac{1}{\chi^2}\, i(\chi)H   \, .
\end{equation}
Because $H$ is antisymmetric, it is obvious that $c$ is orthogonal
to $\chi,$ so that, for the corresponding value $\lambda_c$ of
$\lambda,$ (\ref{standardexpress}) takes the simple form:
\begin{equation}\label{expresswithc}
\lambda_{c\pm} = \pm \frac{\sqrt{\Delta_c}}{\chi^2} \ , \quad
\Delta_c \equiv - \chi^2\,c^2 \ .
\end{equation}
Now, expression (\ref{solutyybot}) for the solution $y,$  $y = c +
\lambda_c \, \chi,$ leads, by product by $\chi$, to $y\cdot\chi =$
$\lambda_c\, \chi^2$. Taking its square and substituting in it the
value (\ref{expresswithc}) of $\lambda_c$, one obtains:
\begin{equation}\label{Deltasubc}
(y \cdot \chi)^2 = \Delta_c  \quad ,
\end{equation}
so that the form (\ref{C-4}) of the coordinate condition
$\textsf{C}$ ensures the strict positiveness of the discriminant
$\Delta_c$. Then, for $y_* = c$ and the values (\ref{expresswithc})
of $\lambda_c$, equations (\ref{solutyybot}) lead, after a little
analysis, to the following result:

%
%
%
\begin{proposition}\label{yplusminusforchiequalxi}
Under the coordinate condition {\rm \textsf{C}}, for non-null
configurations $\chi^2 \neq 0,$ the null propagation equations {\rm
(\ref{ellnullwithy})} admit two real solutions  $y_\pm$  given by
\begin{equation}\label{yforregularconfig}
y_\pm = c \pm |c|\, \nu \, , \qquad \nu \equiv
\frac{\chi^2}{|\chi|^3} \chi \, ,
\end{equation}
where $c$ and $\chi$ are, respectively, given by {\rm (\ref{defc})}
and {\rm (\ref{defvectorchi})}.
\end{proposition}
%
%
%
In the above proposition and in what follows we denote $|v|$ the
modulus of a vector $v$, $|v| \equiv \sqrt{|v^2|}$.

We have seen that, for the particular solution $y_* = c, $ the
discriminant $\Delta$ given by (\ref{standardexpress}) takes the
value $\Delta_c$ given by (\ref{expresswithc}). To what extent the
discriminant $\Delta$, corresponding to other particular solutions
$y_* \neq c ,$ differs from the value $\Delta_c$?

The question is pertinent because we want to unify expressions
(\ref{yunica}) and (\ref{yforregularconfig}) so as to have a single
expression for every coordinate domain, irrespective of the causal
orientation of the configuration vector $\chi$. To answer this, note
that $c$ is the particular solution to the linear system orthogonal
to $\chi$ and that, from (\ref{solutyybot}), any $y_*$ differs from
$c$ by a term of the form $\lambda \chi$. More precisely, $c = y_* -
\frac{y_* \cdot \chi}{\chi^2} \chi$. Then, a straightforward
calculation leads to:
%
%
%
\begin{lemma}\label{DeltaIndptxi}
Under the coordinate condition {\rm \textsf{C}}, for non-null
configurations $\chi^2 \neq 0,$ the discriminant $\Delta$ and each
one of the solutions $y_{\pm}$ of the quadratic equation {\rm
(\ref{firstquadratic})}, given in {\rm (\ref{standardexpress})}, are
independent of the particular solution chosen and, therefore, of the
subsidiary vector $\xi$ too. The discriminant is positive and its
invariant value is
\begin{equation}\label{invariantDelta}
\Delta = \Delta_c = - \chi^2 c^2 > 0   \quad ,
\end{equation}
where $c$ is given by {\rm (\ref{defc}).}
\end{lemma}
%
%
%

\vspace{2mm} \noindent {\textbf{Unified expression for any
configuration.}} As already stated, we want to find, for every
coordinate domain, a {\em sole} expression valid for the {\em whole}
domain, i.e. valid for {\em all} the configurations
$\{\gamma_A(\tau^A)\}$ that could correspond to the events of the
domain.
Nevertheless, in principle, by continuously changing of events in
the coordinate domain, we can make these configurations to have a
vanishing, positive or negative value of $\chi^2$. When this is the
case, \emph{none} of the two {\em standard expressions}
(\ref{standardexpress}) of $\lambda$ for $\chi^2 \neq 0$ comes down
to the sole solution (\ref{solelambda}) for $\chi^2 = 0$ and,
consequently, none of the expressions (\ref{yforregularconfig})
reduces to expression (\ref{yunica}), both of them becoming
degenerate. To correct this degeneration, it is sufficient to
multiply numerator and denominator of the second member of
(\ref{standardexpress}) by the conjugate of the numerator. Once this
is done, taking into  account proposition \ref{expresSolPart} one
has the following result.
%
%
%
\begin{theorem}\label{theotwoanalyticsols}
In all the emission coordinate region $\mathcal{C}$, the solutions to
the null propagation equations {\rm (\ref{ellnullwithy})} are real
and admit the expression:
\begin{equation}\label{yplusminus}
y_\pm = y_* + \lambda_\pm \chi \, , \quad \lambda_\pm \equiv
-\frac{y_*^2}{(\chi \cdot y_*) \pm \sqrt{\Delta}} \, , \quad \Delta
\equiv (\chi\cdot y_*)^2 - \chi^2\,y_*^2 \, ,
\end{equation}
where $y_*$ and $\chi$ are respectively given by {\rm
(\ref{solutyorto})} and {\rm (\ref{defvectorchi})}. For null
configurations $\chi^2 = 0,$ the sole solution is  given by one of
these expressions, the other becoming degenerate.
\end{theorem}
%
%
%


\section{The emission conditions \textsf{E}}
\label{sec-6}
The expressions (\ref{yplusminus}) of theorem
\ref{theotwoanalyticsols} give the Minkowski events $y$ that can be
related to the configuration events $\{\gamma_A(\tau^A)\}$ by means
of emission or reception of null signals. But we want $y$ to be the
events reached by null signals \emph{emitted} from the configuration
events. Thus we have to impose on $y$ the emission conditions
\textsf{E}. This will select a special class of configurations,
which will be called \emph{emission configurations}.

    In a first step we shall impose the intermediate emission-reception
condition \textsf{E-R}, guaranteeing that either all the signals at
$y$ have been emitted by the configuration, or all the signals at
$y$ will be received by the configuration; such a configuration will
be called an \emph{emission-reception configuration}. And in a second
step we shall choose with \textsf{E} the first of these two cases.

As we have seen in section \ref{sec-2}, the emission-reception
condition \textsf{E-R} states that all the null vectors $m_A$
joining $\gamma_A$ and $y$, $m_A \equiv y - \gamma_A,$ must have the
same (past or future) orientation, i.e. $\epsilon \, m_A \cdot m_B <
0\,, \, \forall \, A,B\, . $ In terms of the relative position
vectors $e_a$ of the configuration $\{\gamma_A(\tau^A)\}$ given by
(\ref{defeA}) and (\ref{defe4ea}), one has:
\begin{equation}\label{lsyes}
\begin{array}{ll}
m_4 - m_a = \gamma_a - \gamma_4 = e_a \\
m_a - m_b = \gamma_b - \gamma_a = e_b - e_a \quad ,
\end{array}
\end{equation}
so that, because $(m_A - m_B)^2 = - 2 \, m_A\cdot m_B,$ we have:
%
%
%
\begin{proposition}\label{emissionconfig}
For a regular configuration $\{\gamma_A(\tau^A)\}$ to be an
emission-reception configuration it is necessary and sufficient that
all their relative positions be space-like. In other words: the null
directions $m_A$ verify the emission-reception condition {\rm
\textsf{E-R}} if, and only iff,
\begin{equation}\label{EqSpatialOrient}
{\rm (\textsf{\em E-R})}_4: \quad \epsilon \, (e_a)^2 > 0 \quad ,
\quad \epsilon \, (e_a - e_b)^2 > 0 \quad .
\end{equation}
\end{proposition}
%
%
%
%

Let us note that, in particular, this proposition tells us that, in
the grid $\mathcal{T} \approx$ $\mathbb{R}^4$ of the $\tau$'s, the
domains of emission coordinates $\{\tau^A\}$ are in the interior of
the region determined by ($\ref{EqSpatialOrient}$), so that the
points of the grid $\mathcal{T}$ in the complementary region
certainly have \emph{no} physical meaning.

Equations (\ref{yplusminus}) of theorem \ref{theotwoanalyticsols}
show us that, under the coordinate condition {\rm \textsf{C}}, two
real and definite solutions to the null propagation system may
correspond to every configuration $\{\gamma_A(\tau^A)\}$. Now, under the
additional emission-reception condition {\rm \textsf{E-R}}, one of
these definite solutions may be that of emission and the other of
reception. This last one finally has to be detected and discarded by
means of the emission condition {\rm \textsf{E}}, which, under the
{\rm \textsf{E-R}} one, reduces to:
\begin{equation}\label{emissioncondony}
{\rm \textsf{E}_4:}  \quad    \epsilon  \, y\cdot u < 0  \, ,
\end{equation}
where $u$ is any given, everywhere non vanishing, future-pointing
time-like vector.

For non-null configurations, $\chi^2 \not= 0,$ the general
expression (\ref{yplusminus}) gives the two admissible solutions
under the coordinate condition \textsf{C}. These solutions also
admit the non-unified expression given in proposition
\ref{yplusminusforchiequalxi}. From it we obtain:
\begin{equation} \label{y+doty-}
y_+ \cdot y_- = 2 \, c^2
\end{equation}
Then, (\ref{invariantDelta}) implies that if $\epsilon \, \chi^2 >0$
({\em respectively}, $\epsilon \, \chi^2 < 0$) then  $\epsilon \, c^2 < 0$
({\em respectively}, $\epsilon \, c^2 > 0$ ) and, as a consequence of
(\ref{y+doty-}), $y_+$ is future-pointing iff $y_-$ is future-pointing
({\em respectively}, past-pointing). Thus, we have:
%
%
%
\begin{lemma}\label{y+y-}
If $\chi$ is a time-like vector, $\epsilon \, \chi^2 < 0$, then only
one of the solutions {\rm (\ref{yplusminus})} corresponds to an
emission configuration.

If $\chi$ is a space-like vector, $\epsilon \, \chi^2 > 0$, then the
two solutions of {\rm (\ref{yplusminus})} correspond to either two
emission configurations or two reception configurations.
\end{lemma}
%
%
%
%

When $\chi$ is a time-like vector, can we detect the solution which
corresponds to an emission configuration? The answer is affirmative.
Indeed, as a consequence of lemma \ref{DeltaIndptxi}, each one of
the solutions $y_{\pm}$ is independent of $\xi$. Thus, taking for
them the non-unified expression given in proposition
\ref{yplusminusforchiequalxi}, we obtain (when $\epsilon\, \chi^2 <
0$):
$$
\epsilon \, (\epsilon \chi) \cdot y_{\pm} = \chi \cdot y_{\pm} = \pm
|c|\, |\chi| \, ,
$$
and, consequently, we can state:
%
%
%
\begin{proposition} \label{chi-timelike}
For a space-like emission configuration, $\epsilon \, \chi^2 < 0$,
the sole emission solution is $y_+$ ({\em respectively}, $y_-$) if
$\epsilon \chi$ is past-pointing ({\em respectively},
future-pointing).
\end{proposition}
%
%
%
%

When $\chi$ is a space-like vector, we can know a priori whether the two
solutions $y_{\pm}$ correspond to emission configurations. Indeed,
in this case $c$ is a time-like vector. Then, making use again of
the non-unified expression for $y_{\pm}$ given in proposition
\ref{yplusminusforchiequalxi}, we obtain:
$$
\epsilon \, c \cdot y_{\pm} = \epsilon c^2 < 0 \, ,
$$
and, consequently, we can state:
%
%
%
\begin{proposition} \label{chi-space-like}
For a time-like emission configuration, $\epsilon \, \chi^2
> 0$ the solutions $y_{\pm}$ correspond to emission
({\em respectively}, reception) configurations if $c$ is future-pointing
({\em respectively}, past-pointing).
\end{proposition}
%
%
%
%

Finally, for a null configuration, $\chi^2 = 0,$ proposition
\ref{yparachidosnulo} gives the sole admissible solution under the
coordinate condition \textsf{C}, which of course also reduces to one
of the solutions of the general expression (\ref{yplusminus}). In
this case $\Delta = (y_* \cdot \chi)^2$ and, consequently, the non
degenerate solution in (\ref{yplusminus}) is $y_+$ ({\em
respectively}, $y_-$) if $ \epsilon \, (\epsilon \chi) \cdot y_* =
\chi \cdot y_* > 0$ ({\em respectively}, $<0$). Thus, we can state:
%
%
%
\begin{proposition} \label{chi-light-like}
For a null emission configuration, $\chi^2 = 0$, the non degenerate
solution is $y_+$ ({\em respectively}, $y_-$) if $\epsilon \chi$ is
past-pointing ({\em respectively}, future-pointing).
\end{proposition}
%
%
%
%
It is worth remarking the attachment between the results stated in
propositions \ref{chi-timelike} and \ref{chi-light-like}. If we
continuously change to a null configuration coming from a space-like
one, the configuration vector $\chi$ keeps its future or past
orientation. If $\chi$ is past-pointing ({\em respectively},
future-pointing) we must take the (continuous) expression $y_+$
({\em respectively}, $y_-$) in both the null and space-like regions.


\section{The coordinate
transformation from emission to inertial coordinates} \label{sec-7}
\subsection{Front and back emission coordinate domains}
Theorem \ref{theotwoanalyticsols} shows that the emission coordinate
region ${\cal C}$ is mapped with {\em two} local charts and gives analytical
expressions for the transformation between emission and inertial
coordinates for the two corresponding coordinate domains.

These expressions of the coordinate transformation and the study of
the emission conditions in section \ref{sec-6} show that the causal
character of the configuration of the emitters differs for
the two emission coordinate domains. Space-like and null
configurations only admit one of the solutions given in theorem
\ref{theotwoanalyticsols} as emission solution. And, of course, the
coordinate domain of this solution contains by continuity a
time-like configuration region. The coordinate domain of the other
solution only contains a time-like configuration region.

We see that the two coordinate domains differ enough in the causal
character of their emitter configurations. We shall call {\em front
emission coordinate domain} ${\cal C}^F $ the coordinate domain that
contains events with the three possible causal configurations. More
precisely, ${\cal C}^F $ contains all the space-like configuration
region ${\cal C}_s$, all the null configuration region ${\cal
C}_\ell$ and a part ${\cal C}^F_t$ of the time-like configuration
region ${\cal C}_t$. We shall call {\em back emission coordinate
domain} ${\cal C}^B$ the other coordinate domain, that only contains
events with time-like configurations.\footnote{Let us remember that
it is usual to call {\em coordinate domain} the open set $U$ of any
local chart $(U, \phi)$ of the atlas defining a differentiable
manifold. This appellation requires attention because a coordinate
domain is not necessarily a domain, but simply a (not necessarily
connected) topological open set. In fact, meanwhile the front
emission coordinate domain ${\cal C}^F $ is generically connected,
the back emission coordinate domain ${\cal C}^B$ is generically the
disjoint union of four connected components \cite{Pozo-Escola}.} In
fact, ${\cal C}^B$ $= {\cal C}_t - {\cal C}^F_t$.

Moreover, the two time-like regions have the same codomain in the
grid, namely, $\Theta({\cal C}^F_t)$  $= \Theta({\cal C}^B)$. This
relation is very important. It says that, whatever be the four
values $\{\tau^A\}$ received by a user in the coordinate region
${\cal C}^F_t$  (resp. ${\cal C}^B$), another user in the coordinate
region ${\cal C}^B$ (resp. ${\cal C}^F_t$) may receive the {\em
same} values $\{\tau^A\}$. In other words: a user that only receives
four times $\{\tau^A\}$ defining a time-like configuration of the
emitters is {\em unable} to detect in what part of the region ${\cal
C}_t$, in ${\cal C}^F_t$ or in ${\cal C}^B$, he is.
\subsection{Orientation of a positioning system}
To be able to detect in what of these regions a user is, a notion of
orientation of a positioning system is necessary. We give the
following one:

\begin{definition}
The {\em orientation} of a positioning system with respect to an
event of its emission coordinate region ${\cal C}$ is the sign
$\hat{\epsilon}$ of the coordinate condition scalar:
 \begin{equation}
 \label{deforientation}
\hat{\epsilon} \equiv \textit{sgn} * (m_1 \wedge m_2 \wedge m_3
\wedge m_4) \quad .
 \end{equation}
 \end{definition}

We have seen in section \ref{sec-2-2} that, in the emission region
${\cal R}$, the hypersurface $\cal J$ separates the emission
coordinate region $\cal C$ in two open sets. We can now identify
them with the above front and back coordinate domains ${\cal C}^F$
and ${\cal C}^B$, respectively. Because $\cal J$ is the hypersurface
where the coordinate condition $\textsf{C}$ is not verified
(vanishing Jacobian $j_\Theta(x) = 0$), the non vanishing member of
its inequality has a constant sign in every coordinate domain ${\cal
C}^F$ and ${\cal C}^B$, so that we have the simple but important
result:
 \begin{proposition}
 \label{epsilonconstant}
The orientation $\hat{\epsilon}$ of a positioning system is constant
in every one of the coordinate domains ${\cal C}^F$ and ${\cal
C}^B$.
 \end{proposition}

The same way that leads to the form (\ref{C-4}) of the coordinate
condition $\textsf{C}$ shows that $\hat{\epsilon} = \textit{sgn}\,\,
(y\cdot\chi)$. Because $y$ is necessarily future-pointing for a
emitted signal, if $\chi$ is time-like or null we can be sure that
the sign of $(y\cdot\chi)$ is the same as that of the sign of
$(u\cdot \chi)$ for any everywhere non vanishing future-pointing
time-like vector $u$. This last sign is plus or minus according to
the past- or future-pointing character of $\epsilon\chi$. Taking
into account propositions \ref{chi-timelike} and
\ref{chi-light-like}, one has the following:

 \begin{proposition}
 \label{hatepsilonuchi}
In the regions ${\cal C}_s$ and ${\cal C}_\ell$, the orientation
$\hat{\epsilon}$ of a positioning system is given by
 \begin{equation}
 \label{ucdotchi}
 \hat{\epsilon} = \textit{sgn}\,(u\cdot\chi)
 \end{equation}
 for any future-pointing time-like vector $u$.
\end{proposition}

\subsection{Explicit expression of the coordinate
transformation from emission to inertial coordinates}
We shall comment below in section \ref{sec-8} that for users in the
region ${\cal C}_t$ the determination of the orientation needs
additional information. But for the moment, Theorem
\ref{theotwoanalyticsols} and the above propositions lead us to the
following result.
\begin{theorem}
\label{teotransffinal} Let $\gamma_A$ be the position vectors of the
world-line equations $\gamma_A(\tau^A)$ of the four emitters of a
positioning system with respect to an inertial coordinate system
$\{x^\alpha\}$, and $\{\tau^A\}$ their emission coordinates. In all
the emission coordinate region ${\cal C}$ $= {\cal C}^F \cup {\cal
C}^B$, the coordinate transformation $x = \kappa(\tau^A)$ is given
by:
\begin{equation}
\label{generaltransf} x = \gamma_4 + y_* - \frac{y_*^2 \,
\chi}{(y_*\cdot \chi ) + \hat{\epsilon}\sqrt{(y_*\cdot \chi )^2 -
y_*^2 \chi^2}}
\end{equation}
where $y_*$ is the quantity given by {\rm (\ref{solutyorto})},
$\chi$ is the configuration vector {\rm (\ref{defvectorchi})} and
$\hat{\epsilon}$ is the orientation {\rm (\ref{deforientation})} of
the positioning system with respect to the event that receives the
data $\{\tau^A\}$.
\end{theorem}

This is the main result reached in this paper. It rest to analyze
now to what extent this expression (\ref{generaltransf}) may be
determined by a user from the data received by him.

\section{The central
region of a positioning system. Computational and observational
orientation} \label{sec-8}

The world-lines $\gamma_A(\tau^A)$ of the emitters in an inertial
system $\{x^\alpha\}$, as well as the space-time metric in it, are
here supposed known `background' data for any user of the
positioning system (the world-lines can be pre-determined initially
or broadcast in real time). Thus, any user who receives only the
emission data $\{\tau^A\}$ is able to compute the quantities
$\gamma_4$, $y_*$, $\chi$ appearing in (\ref{generaltransf}). He has
to follow the four steps:

\emph{Step 1.} Compute the four position vectors of the emitters,
$\gamma_A = {\rm {\small O}}\gamma_A(\tau^A)$, for the received
values $\{\tau^A\}$.

\emph{Step 2.} Choose a reference emitter, say $\gamma_4$, and
compute the position vector of the three referred emitters $e_a =
\gamma_a(\tau^a) - \gamma_4(\tau^4)$.

\emph{Step 3.} Compute the configuration scalars $\Omega_a \equiv
\frac{1}{2}e_a^2$, the configuration vector $\chi \equiv *(e_1\wedge
e_2\wedge e_3)$, and the configuration bivectors $E^a \equiv
*(e_{a+1} \wedge e_{a+2})$ and $H \equiv \Omega_aE^a$.

\emph{Step 4.} Choose a transversal vector $\xi$, $\xi\cdot \chi
\neq 0$, and compute $y_* \equiv \frac{1}{\xi\cdot \chi} i(\xi)H$.

At this level, the user has computed the quantities $\gamma_4$,
$y_*$, $\chi$. But he is also able to compute in what of the
coordinate regions, ${\cal C}_t$, ${\cal C}_\ell$ or ${\cal C}_s$ of
the positioning system he is; for this, one additional step is
sufficient:

\emph{Step 5.} Determine the sign of $\epsilon \chi^2$.

\noindent Then, according to their definitions, the user is in
${\cal C}_t$, ${\cal C}_\ell$ or ${\cal C}_s$ if this sign of
$\epsilon \chi^2$ is $> 0$, $= 0$ or $< 0$ respectively.

Now, suppose that the user is in ${\cal C}_\ell$ or ${\cal C}_s$.
Another step allows him to compute the orientation $\hat\epsilon$:

\emph{Step 6.} Determine the sign of $u\cdot \chi$ for a
future-pointing time-like vector $u$, arbitrarily chosen.

\noindent Then, according to proposition \ref{hatepsilonuchi},
$\hat{\epsilon} = \textit{sgn}\,(u\cdot\chi)$.

We shall call {\it central region} of a positioning system the
region ${\cal C}^C$ $\equiv {\cal C}_s \cup {\cal C}_\ell$. A part
of theorem \ref{teotransffinal} may be then stated as follows.

\begin{proposition}
Let $\gamma_A$ be the position vectors of the known world-line
equations $\gamma_A(\tau^A)$ of the four emitters with respect to an inertial
coordinate system. Then the users of the central region ${\cal C}^C$ of the
positioning system can obtain their position $\{x^\alpha\}$ in the inertial
system by {\em computation} from their sole emission coordinates
$\{\tau^A\}$.
\end{proposition}

What about the users in ${\cal C}_t$, i.e. out of the central region
${\cal C}^C$? We have seen that everywhere $\hat{\epsilon} =
\textit{sgn}\,\, (y\cdot\chi)$. Nevertheless, the null vector $y$
involves not only the initial data $\gamma_A(\tau^A)$ and the
received data ${\tau^A}$, but is the solution we are looking for, so
that the quantity $y\cdot\chi$ cannot be computed before itself. For
this reason proposition \ref{hatepsilonuchi} is {\em exclusive}:
{\em only} the users of the central region ${\cal C}^C$ are able to
{\em compute} the orientation $\hat\epsilon$ of the positioning
system.

Can the users in ${\cal C}_t$ know the orientation of the
positioning system? It is possible to show that all the users of the
coordinate region ${\cal C}$, be them in the central region ${\cal
C}^C$ or not, can determine the orientation of the positioning
system if, in addition to the reception of the $\{\tau^A\}$, they
are able to {\em observe} the emitters in their celestial sphere.
But this fact will be analyzed in a forthcoming paper.


\section{Discussion and work in progress}
\label{sec:discussion}

In this paper we have obtained the coordinate transformation
(\ref{generaltransf}) between emission coordinates and inertial
coordinates in Minkowski space-time. We use the intrinsic vector
formalism to express the position vector, $x \equiv (x^{\alpha})$,
of every event in the emitter coordinate region as a function $f$ of
the emitter world-lines $\gamma_A(\tau^A):$ $x =$
$f({\rm {\small O}}\gamma_A(\tau^A))$ $\equiv \kappa(\tau^A)$.

This general and compact expression of the coordinate transformation
will be a powerful tool for subsequent applications. For example, we
can particularize it for different choices of the emitter
world-lines which model specific physical situations. In doing so, a
previous basic task appears to be convenient in many cases: to write
our covariant expressions in a 3+1 formalism with respect to an
arbitrary inertial observer, which is the goal of another work
\cite{cfm-min-b}. Moreover, from the expression of the coordinate
transformation we can easily obtain the components of the metric
tensor in emission coordinates and we can study the region where the
emission coordinates are more efficient than the inertial ones
\cite{cfm-min-d}.

In section \ref{sec-6} we have studied the emission conditions in
order to distinguish between emission and reception or mixed
configurations. These results are a necessary tool to carry out
more in-depth analysis of the domains and co-domains of the emission
coordinates. This task will be tackled in another paper
\cite{cfm-min-c} where we will also study the geometry of the
emitter configurations attending to their different causal
character, and we will analyze how this geometry influences the
solutions to the null propagation equations. Some preliminary
results on this question have been presented in \cite{NosERE08}.

The orientation $\hat{\epsilon}$ of a positioning system with
respect to an event (see section \ref{sec-7}) is a concept which has
allowed us to give an explicit expression of the coordinate
transformation in theorem \ref{teotransffinal}. In section
\ref{sec-8} we have pointed out that this orientation
$\hat{\epsilon}$ may be computed from the emission data $\{\tau^A\}$
in the so-called central region. Nevertheless, out of this region,
the determination of $\hat{\epsilon}$ demands an observational
method which will be analyzed elsewhere.

The development of all this theoretical work paves the way to the
study of more realistic, gravitationally influenced, positioning
systems. For example, those defined by emitter world-lines modeling
a satellite constellation around the Earth in a weak gravitational
field. It is worth remarking that for the use and smooth running of
a positioning system one needs, not only the coordinate
transformation to emission coordinates but also a good understanding
of the domains in the space-time and co-domains in the grid. The
analysis of the degenerate configuration data, $\chi = 0$, and of
the hypersurfaces of the emission region ${\cal R}$ where the
Jacobian $j_{\Theta}(x)$ vanishes must also be well understood. We
already know that the events of vanishing Jacobian are, and only
are, those for which any user in them can see the four emitters on a
circle in his celestial sphere \cite{Pozo-Escola}. This includes the
possibility for the user seeing less than four satellites, when some
of them are in the shadows of the others.

Obviously, in realistic situations, the inertial coordinate system
considered here must give rise to more useful ones. Namely: the
International Celestial Reference System (ICRS) for the positioning
system based on four millisecond pulsars, valid for the Solar
System, as proposed in \cite{Galactic}; the Barycentric Celestial
Reference System (BCRS), to compare the planet trajectories (Earth
at least and already) obtained in this millisecond pulsar system
with the ones obtained by standard astronomical observations; the
Geocentric Celestial Reference System (GCRS) or related World
Geodetic System 84 (WGS84) or International Earth Reference System
(ITRS) for the applications of relativistic positioning systems to
the Global Navigation Satellite Systems (GNSS).

\ack This work has been supported by the Spanish Ministerio de
Educaci\'on y Ciencia, MEC-FEDER project FIS2006-06062 and Spanish
Ministerio de Ciencia e Innovaci\'on, MICIN-FEDER project
FIS2009-07705.

\appendix

\section{Proof of proposition \ref{expresSolPart}}
\label{A-SolCov}

From lemma \ref{Y=chi}, to prove proposition \ref{expresSolPart} we
must obtain a particular solution of the main linear system
(\ref{linearsys}).
More precisely, we must obtain the particular solution $y_*$ which
is orthogonal to a chosen transversal vector $\xi$, $ \xi \cdot \chi
\neq 0$.

In order to obtain such a solution in a covariant way we begin by
studying a regular linear system $\phi_A \cdot z = \Omega_A$. In
this case, the vectors $\{\phi_A\}$ define a base, that is, $\phi_1
\wedge \phi_2 \wedge \phi_3 \wedge \phi_4 \not = 0$. Then, the sole
solution to the system takes the expression $ z = \Omega_A\,
\phi^A$, $\{\phi^A\}$ being the dual base. More explicitly, we have:
\begin{lemma} \label{A-lemma-RLS}
The solution $z$ to the regular linear system
\begin{equation}  \label{A-RLS-1}
    \phi_A \cdot z = \Omega_A \; , \qquad A= 1,2,3,4 \ ,
\end{equation}
is given by
\begin{equation}   \label{A-RLS-2}
    z = \Omega_A\, \phi^A \ .
\end{equation}
where
\begin{equation}   \label{A-RLS-3}
\phi^A \equiv \frac{1}{3! D} \epsilon^{APQR} *(\phi_P \wedge \phi_Q
\wedge \phi_R) \ , \quad D \equiv *(\phi_1 \wedge \phi_2 \wedge
\phi_3 \wedge \phi_4)  \ .
\end{equation}
\end{lemma}

Now, we regularize the main linear system (\ref{linearsys}) by
adding a new equation as follows. Given a vector $\xi$ such that
$\xi \cdot \chi \not=0$, let us consider a linear system of the form
(\ref{A-RLS-1}) for the unknown $z$, with
\begin{eqnarray}  \label{A-PS-1}
    \phi_A = e_a \ , \qquad \Omega_A = \Omega_a \ , \quad A = 1,2,3 \ ;\\
    \phi_4 = \xi \ , \qquad \ \ \Omega_4 = 0 \ .
\end{eqnarray}
This linear system is regular since:
\begin{equation}  \label{A-PS-2}
\hspace{-5mm}
\begin{array}{lll}
D & \equiv & *(\phi_1 \wedge \phi_2 \wedge
\phi_3 \wedge \phi_4)
   = *(e_1 \wedge e_2 \wedge e_3 \wedge \xi)\\[1mm]
& = & - *(\xi \wedge e_1 \wedge e_2 \wedge e_3) = - i(\xi)*(e_1
\wedge e_2 \wedge e_3) = - \xi \cdot \chi \not=0 \ .
\end{array}
\end{equation}
Consequently, the sole solution to this system may be obtained as
stated in lemma \ref{A-lemma-RLS}. Now $\Omega_4 = 0$, and then we
only need to calculate the vectors $\phi_a, \ a=1,2,3,$ of the dual
basis given in (\ref{A-RLS-3}). They take, in this case, the
expression:
\begin{equation}  \label{A-PS-4}
\hspace{-5mm}
\begin{array}{lcl}
\phi^a & \equiv & \frac{1}{3! D} \epsilon^{aPQR} *(\phi_P \wedge
\phi_Q \wedge \phi_R) = \frac{1}{2 \, D} \epsilon^{abc4} * (e_b
\wedge e_c  \wedge \xi)  \\[2mm]
& = & \frac{1}{2 \, D} \epsilon^{abc} * (\xi \wedge e_b \wedge e_c)
 = - \frac{1}{2 \, D} \epsilon^{abc} i(\xi)* (e_b \wedge e_c)  =
\frac{1}{ \xi \cdot \chi}i(\xi)E^a \ ,
\end{array}
\end{equation}
where
\begin{equation}\label{defEaBis}
    E^a \equiv  *(e_{a+1} \wedge e_{a+2}) \ .
\end{equation}
Then, the solution (\ref{A-RLS-2}) becomes now
\begin{equation}   \label{A-PS-6}
    z = \frac{1}{\xi\cdot\chi} \ i(\xi)H \ , \qquad H
\equiv  \Omega_a E^a \ .
\end{equation}
Finally, note that this vector $z$ is a solution to the main linear
system and it is orthogonal to $\xi$. Thus it is the particular
solution $y_{*}$ that we are looking for.

\end{document}